\documentclass[referee]{chjaa}          

\usepackage{graphicx,times}             

\begin{document}

   \title{Rotation profiles of solar-like stars with magnetic fields}

   \volnopage{Vol.0 (200x) No.0, 000--000}      
   \setcounter{page}{1}           

   \author{Wuming Yang
      \inst{1}
   \and Shaolan Bi
   \inst{2, 3}
      }

\institute{Department of Physics and Chemistry, Henan Polytechnic
University, Jiaozuo 454000, PR China. wuming.yang@hotmail.com\\
\email{yangwuming@ynao.ac.cn}
 \and
    Department of Astronomy, Beijing Normal University,
    Beijing 100875, China. bisl@bnu.edu.cn\\
 \and
    National Astronomical Observatories/Yunnan Observatory,
    Chinese Academy of Sciences, Kunming 650011, China\\
          }


\abstract{The aim of this work is to investigate rotation profile of
solar-like stars with magnetic fields. A diffusion coefficient of
magnetic angular momentum transport is deduced. Rotating stellar
models with different mass are computed under the effect of the
coefficient. Then rotation profiles are obtained from the
theoretical stellar models. The total angular momentum of solar
model with only hydrodynamic instabilities is about 13 times larger
than that of the Sun at the age of the Sun, and this model can not
reproduce quasi-solid rotation in the radiative region. However, not
only can the solar model with magnetic fields reproduce an almost
uniform rotation in the radiative region, but its total angular
momentum is consistent with helioseismic result at the level of 3
$\sigma$ at the age of the Sun. The rotation of solar-like stars
with magnetic fields is almost uniform in the radiative region. But
there is an obvious transition region of angular velocity between
the convective core and the radiative region of models with 1.2 -
1.5 $M_{\odot}$, where angular velocity has a sharp radial change,
which is different from the rotation profile of the Sun and massive
stars with magnetic fields. Moreover the changes of the angular
velocity in the transition region increase with the increasing in
the age and mass. \keywords{stars: evolution -- stars: rotation --
stars: magnetic fields } }

\authorrunning{Yang, W. and Bi, S. }
\titlerunning{Rotation profiles of solar-like stars}

\maketitle

\section{Introduction}
Helioseismology has given us detailed internal information about the
structure and rotation of the Sun: the Sun's rotation is slow in the
core and is almost uniform in the radiative region, but the angular
velocity has a latitudinal gradient in the convective zone (Gough et
al. \cite{Gough96}; Schou et al. \cite{scho98}; Chaplin et al.
\cite{chap99}). Although the data of inversion of solar-like stars
are limited, it has been revealed that the localized information of
stellar interior can be given by asteroseismology (Gough \&
Kosovichev \cite{gough93}; Gough \cite{gough98}; Roxburgh et al.
\cite{rox98}; Berthomieu et al. \cite{bert01}; Basu et al.
\cite{basu02}; Basu \cite{ basu03}). Furthermore, it has been shown
that localized information on the internal rotation profile of a
solar-like star can be obtained from the frequencies of oscillations
(Gough \& Kosovichev \cite{gough93}; Gough \cite{gough98}; Goupil et
al. \cite{goup96}; Lochard et al. \cite{loch04, loch05}). The
information on the internal rotation of $\beta$ Cepheid has already
been provided by asteroseismology (Aerts et al. \cite{aert03}). And
using the data of the Microvariability and Oscillation of Star
(MOST) satellite, Walker et al. (\cite{walk07}) found that kappa1
Ceti has a differential rotation profile closely resembling that for
the Sun. With ongoing and forthcoming space seismic missions:
COvection, ROtatin and planetary Transits (COROT) (Baglin
\cite{bagl06}) and Kepler (Christensen-Dalsgaard \cite{chri07}), it
is possible to extract the information on the internal rotation
profile of solar-like stars.

Moreover, magnetic fields of the active regions on solar surface are
believed to originate from strong toroidal magnetic fields generated
by solar dynamo at the base of the convective zone. The
understanding of both the stellar magnetic activity and the
generation of magnetic fields is dependent on the information about
the interior rotational properties of stars (Thompson et al.
\cite{thom03}; Fan \cite{fan04}; Charbonneau \cite{char05}). However
the evolution of rotation profile inside stars is poorly understood.
Therefore it is an important problem to get a global picture of the
evolution of rotation profile inside stars.

The influence of rotation on the stellar structure and evolution is
studied by many investigators (Kippenhaln \& Thomas \cite{kipp70};
Endal \& Sofia \cite{endal76}; Pinsonneault et al. \cite{pins89};
Meynet \& Maeder \cite{meyn97}; Huang et al. \cite{huang07}).
Redistribution of angular momentum within the interiors of stars has
also been considered by many authors (Endal \& Sofia \cite{endal81};
Chaboyer et al. \cite{chab95}; Maeder \& Meynet \cite{maed00};
Palacios et al. \cite{pala03}; Huang \cite{huang04}). These studies
show that hydrodynamic angular momentum transport processes are
unefficient in stars. Therefore magnetic angular momentum transport
or other mechanisms should be considered in rotating stars. In
addition, magnetic angular momentum transport in massive stars has
been investigated by Maeder \& Meynet (\cite{maed03, maed04,
maed05}). The massive stars with magnetic fields rotate almost as a
solid body throughout the whole star (Maeder \& Meynet
\cite{maed04}). Eggenberger et al. (\cite{egge05}) and Yang \& Bi
(\cite{yang06}) study the rotation profile of the Sun and show that
the quasi-solid rotation in the Sun can be achieved by considering
the effect of the magnetic fields. In this paper, we mainly focus on
the internal rotation profiles of solar-like stars. In Sect. 2
diffusion coefficient of magnetic angular momentum transport is
given. In Sect. 3 numerical calculation and results are presented.
Then, discussion and conclusion are made in Sect. 4.

\section{Diffusion coefficient of magnetic angular momentum transport}

Spruit (\cite{spr99}; \cite{spr02}) developed the Tayler-Spruit
dynamo, which can generate magnetic fields in the radiative region
of differentially rotating stars. These fields are predominantly
azimuthal components, $B\sim B_{\phi}$. If magnetic fields exist in
stars, magnetic angular momentum transport can be described by
magnetic induction and momentum equations. For a constant magnetic
diffusivity and shellular rotation (Zahn \cite{zahn92}), under
axisymmetry and only considering Lorentz force, the azimuthal
components of the induction and momentum equations are (Barnes et
al. \cite{barn99}; Yang \& Bi \cite{yang06})
\begin{equation}
  \frac{\partial B_{\phi}}{\partial t}+\eta(\frac{1}{r^{2}\sin^{2}\theta}
  -\nabla^{2})B_{\phi}=r\sin\theta \mathbf{B_{p}}\cdot\nabla\Omega \,,
  \label{induc1}
\end{equation}
\begin{equation}
  \rho r^{2}\sin^{2}\theta\frac{\partial \Omega}{\partial t}=
  \frac{1}{4\pi}\mathbf{B_{p}}\cdot \nabla (r\sin\theta B_{\phi}) \,.
  \label{moment1}
\end{equation}
If the effect of the magnetic diffusivity is to limit the growth of
the toroidal field after some time, the growth of the instability is
halted by dissipative processes that operate on a timescale $\tau$.
Accordingly the second term on the left-hand side of Eq.
(\ref{induc1}) may be replaced simply by $B_{\phi}/\tau$ (Barnes et
al. \cite{barn99}). Substituting for the second term of Eq.
(\ref{induc1}) and differentiating the Eq. (\ref{induc1}) with
respect to time, one can obtain (Barnes et al. \cite{barn99})
\begin{equation}
  \frac{\partial^{2}B_{\phi}}{\partial t^{2}}+\frac{1}{\tau}
  \frac{\partial B_{\phi}}{\partial t}= r\sin\theta \mathbf{B_{p}}\cdot
  \nabla\frac{\partial \Omega}{\partial t}\,.
\end{equation}
For much longer times than the timescale of the instability, one
would expect the term involving the first time derivative to
dominate (Barnes et al. \cite{barn99}), so that
\begin{equation}
  \frac{1}{\tau}\frac{\partial B_{\phi}}{\partial t} \approx
  r\sin\theta \mathbf{B_{p}}\cdot\nabla\frac{\partial \Omega}{\partial t}\,.
  \label{}
\end{equation}
For shellular rotation $\Omega(r, \theta) \sim\Omega(r)$ (Zahn
\cite{zahn92}), then
\begin{equation}
  B_{\phi}\sim \tau r\sin\theta B_{r} \frac{\partial \Omega}{\partial r}\,.
  \label{bphi}
\end{equation}
Using Eq. (\ref{bphi}), equation (\ref{moment1}) can be rewritten as
\begin{equation}
\begin{array}{ll}
  \rho r^{2}\frac{\partial\Omega}{\partial t}&\approx \frac{1}{4\pi
  \sin^{2}\theta}\mathbf{B_{p}}\cdot \nabla (\tau r^{2}\sin^{2}\theta B_{r}
  \frac{\partial \Omega}{\partial r}) \\
   & \approx \frac{B_{r}}{4\pi }\frac{\partial}{\partial r}
   (\tau r^{2}B_{r}\frac{\partial \Omega}{\partial r}) \\
   & \approx \frac{1}{r^{2}}\frac{\partial}{\partial r}
   (\frac{\tau B_{r}^{2}}{4\pi\rho}r^{4}\rho
   \frac{\partial\Omega}{\partial r})\,.
\end{array}
  \label{jdiff}
\end{equation}
The diffusion coefficient for angular momentum transport can thus be
obtained as
\begin{equation}
  D_{m} =\frac{\tau B_{r}^{2}}{4\pi \rho}.
  \label{dm1}
\end{equation}
In paper I (Yang \& Bi \cite{yang06}) we also got a similar
diffusion coefficient, but it was only an assumption that the
coefficient can be used in the equation of the transport of angular
momentum. From Eq. (\ref{jdiff}) it can be found that the magnetic
angular momentum transport approximately obeys the diffusion
coefficient $D_{m}$.

For a steady equilibrium, the dissipating timescale $\tau$ has to
match the growth timescale $\sigma^{-1}$ of the instability. Using
the growth time scale of magnetic instability given by Pitts \&
Tayler (\cite{pitt85}) and Spruit (\cite{spr99}),
\begin{equation}
  \sigma^{-1}=\frac{\Omega}{\omega_{A}^{2}}, \omega_{A}= \frac{B_{\phi}}{(4\pi
  \rho)^{1/2}r},
\end{equation}
one can get the diffusion coefficient
\begin{equation}
\begin{array}{lll}
  D_{m}&=&\frac{B_{r}^{2}}{4\pi \rho}\frac{\Omega}{\omega_{A}^{2}}\\
  &= &r^{2}\Omega (\frac{B_{r}}{B_{\phi}})^{2}.
\end{array}
\label{dm2}
\end{equation}
Equation (\ref{dm2}) can also be rewritten as
\begin{equation}
  D_{m}=r^{2}\Omega (\frac{\omega_{rA}}{\omega_{A}})^{2},
  \label{dm3}
\end{equation}
where
\begin{equation}
   \omega_{rA}= \frac{B_{r}}{(4\pi \rho)^{1/2}r}.
\end{equation}
Equation (\ref{dm3}) hints that magnetic angular momentum transport
is related to Alfv\'{e}n waves.

The distribution of magnetic fields inside a star is poorly known.
One of the distributions of magnetic fields was given by Spruit
(\cite{spr02})
\begin{equation}
  \frac{B_{r}}{B_{\phi}}=q(\frac{\Omega}{N_{\mu}})^{2},
  \label{bb1}
\end{equation}
where $q = -\frac{\partial \mathrm{ln}\Omega}{\partial
\mathrm{lnr}}$, for the case 0 that the effect of thermal diffusion
can be neglected, namely $N_{\mu}>N_{T}$, and
\begin{equation}
  \frac{B_{r}}{B_{\phi}}=2^{1/4}(\frac{\Omega}{N_{T}})^{1/4}
  (\frac{\kappa}{r^{2}N_{T}})^{1/4}
  \label{bb2}
\end{equation}
for the case 1 with the effect of thermal diffusion. Using the
expressions (\ref{bb1}) and (\ref{bb2}), one can rewrite Eq.
(\ref{dm2}) as
\begin{equation}
  D_{m0}=r^{2}\Omega q^{2}(\frac{\Omega}{N_{\mu}})^{4}
  \label{dmc0}
\end{equation}
for the case 0 and
\begin{equation}
  D_{m1}=2^{1/2}r^{2}\Omega(\frac{\Omega}{N_{T}})^{1/2}
  (\frac{\kappa}{r^{2}N_{T}})^{1/2}
  \label{dmc1}
\end{equation}
for the case 1. Equations (\ref{dmc0}) and  (\ref{dmc1}) are
consistent with the effective magnetic viscosity defined by Spruit
(\cite{spr02}) and Maeder \& Meynet (\cite{maed04}) for the radial
transport of angular momentum. The expression of (\ref{dmc0}) and
(\ref{dmc1}) is only one of the cases of $D_{m}$. Braithwaite
(\cite{brai06}) validates the Tayler-Spruit dynamo scenario, but
which is contrary to the findings of Zahn et al. (\cite{zahn07}).
The rotation profile of massive stars with magnetic fields was
investigated by Maeder \& Meynet \cite{maed04}. And the rotation
profile of the Sun with magnetic fields was studied by Eggenberger
et al. \cite{egge05}. In this work we focus on solar-like stars with
mass 1.0 - 1.5 $M_{\odot}$.

\section{Numerical calculation and results}
\subsection{Angular momentum transport and loss}
The Yale Rotation Evolution Code (YREC7) is used to construct
stellar models in its rotating configuration (Pinsonneault et al.
\cite{pins89}; Guenther et al. \cite{gue92}). All models are evolved
from fully convective pre-main sequence (PMS) to somewhere near the
end of the Main Sequence (MS). The newest OPAL EOS-2005\footnote
{http://physci.llnl.gov/Research/OPAL/} (Rogers \& Nayfonov
\cite{rog02}), OPAL opacity (Iglesias \& Rogers \cite{igl96}), and
the Alexander \& Ferguson (\cite{ale94}) opacity for low temperature
are used. The models take into account diffusion of helium and
metals, using the prescription of Thoul et al. (\cite{tho94}). The
initial chemical composition of the models is fixed at $Z_{0}=0.02$,
$X_{0}=0.706$.

Hydrodynamic instabilities considered in the YREC7 have been
presented by Pinsonneault et al. (\cite{pins89}). It is assumed that
convection enforces solid-body rotation in the convective regions of
a star. Therefore the rotational instabilities are effective only in
the radiative regions. The transport of angular momentum is treated
as (Endal \& Sofia \cite{endal78}; Pinsonneault et al.
\cite{pins89})
\begin{equation}
  \rho r^{2}\frac{\partial \Omega}{\partial t}
  =f_{\Omega}\frac{1}{r^{2}}\frac{\partial}{\partial}
  (\rho r^{4}D\frac{\partial \Omega}{\partial r})
\end{equation}
in the radiative regions of a star, where $f_{\Omega}$ is an
adjustable parameter introduced to represent some inherent
uncertainties in the diffusion equation. In stars with $M \leq 1.5
M_{\odot}$, angular momentum loss due to magnetic braking is treated
as a parameterized formula (Kawaler \cite{kaw88})
\begin{equation}
  \frac{d J}{d t}
  =f_{K}K_{\Omega}(\frac{R}{R_{\odot}})^{1/2}
  (\frac{M}{M_{\odot}})^{-1/2}\Omega^{3}
  \label{agloss}
\end{equation}
to reproduce the Skumanich relationship (Skumanich \cite{sku72}),
where $K_{\Omega} \simeq 1.13\times10^{47}$ g cm$^{2}$ s, $f_{K}$ is
an adjustable parameter. It is assumed that the magnetic braking has
an effect on the whole convective envelope. In some case, a PMS star
could be locked by the surrounding accretion disk. However the disk
can extract angular momentum from the star as well as can supply
angular momentum to the star (Stassun \& Terndrup \cite{stas03}).
Moreover, it has been argued by Matt \& Pudritz (\cite{matta,
mattb}) that the spin-down of PMS stars may not be due to a magnetic
star-disk interaction, but may result from a magnetic stellar wind.
Thus for simplicity, we do not consider the magnetic star-disk
interaction.

The initial angular momentum of a star is still uncertain. Kawaler
(\cite{kaw87}) shows that the angular momentum $J$ of stars more
massive than 1.5 $\mathrm{M}_{\odot}$ is proportional to squared
mass $\mathrm{M}^{2}$. But the mass-momentum relation of stars below
1.5 $\mathrm{M}_{\odot}$ is uncertain. As a first test, we take the
initial angular momentum to be a free parameter. The adjustable
parameters mentioned above are listed in Table \ref{tab1}.

   \begin{table}
   \caption[]{Model parameters}
   \label{tab1}
   \centering
   \begin{tabular}{c c c c c c}
   \hline\hline
     model &  mass  & $J^{a}_{0}$ &  $f_{k}$ & $f_{\Omega1}^{\mathrm{b}}$
           & $f_{\Omega2}^{\mathrm{c}}$\\
           & M$_{\odot}$  & $10^{50}$ g $\mathrm{cm^{2} s^{-1}}$ &    &  & \\
   \hline
    M1.0a&1.0   & 1.591  & 3.0  &  1.0 & 0\\
    M1.0b&1.0   & 1.591  & 3.0  & 1.0  & 0.01 \\
    M1.2&1.2    &1.9095  & 3.0  & 1.0  &0.01 \\
    M1.4&1.4    &1.534   & 1.0  & 1.0  &0.01 \\
    M1.5&1.5    &1.095   & 1.0  & 1.0  &0.01 \\
   \hline
   \end{tabular}
   \begin{list}{}{}
     \item [$^{\mathrm{a}}$] The initial angular momentum;
     \item [$^{\mathrm{b}}$] The value of $f_{\Omega}$ for the coefficient of
     hydrodynamic instabilities (Pinsonneault et al. \cite{pins89});
     \item [$^{\mathrm{c}}$] The value of $f_{\Omega}$ for $D_{m}$; model M1.0a
     with only hydrodynamic instabilities.
   \end{list}
   \end{table}

\subsection{Results of calculation}

   \begin{figure}
     \centering
     \includegraphics[angle=-90, width=6cm]{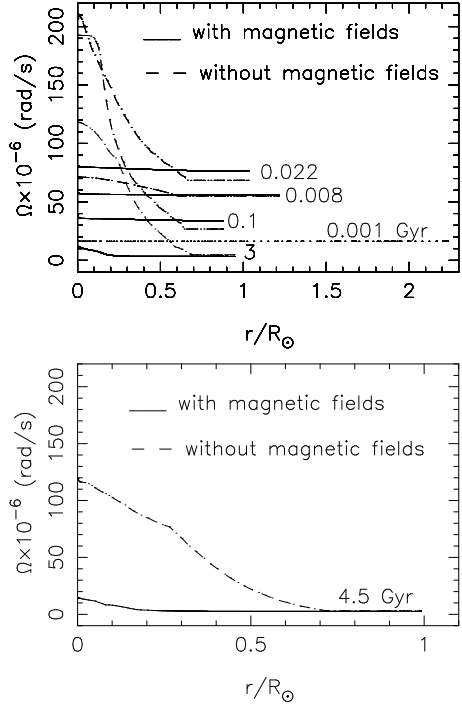}
       \caption{Rotation profiles as a function of radius for
       1.0 $\mathrm{M}_{\odot}$ model at different ages labeled by Gyr.
       The initial angular momentum $J_{0}=1.591 \times 10^{50}$
       g $\mathrm{cm^{2} s^{-1}}$. The dotted line indicates the
       solid-body rotation of the models with and without magnetic
       fields on the pre-main sequence.
        }
       \label{fig1}
   \end{figure}

Figure \ref{fig1} compares the evolution of the internal rotation
profile of 1.0 M$_{\odot}$ models with only hydrodynamic
instabilities given by Pinsonneault et al. (\cite{pins89}) and with
both the hydrodynamic instabilities and magnetic fields. Both models
are evolved from PMS with initial angular momentum $J_{0}=1.591
\times 10^{50}$ g cm$^{2}$ s$^{-1}$ to the age of 4.5 Gyr. During
the PMS phase, although with angular momentum loss from the surface
of models, the rotation rate rapidly increases due to quickly
contracting. The internal rotation profile of model with magnetic
fields has been different from that of model with only hydrodynamic
instabilities when the models are near the zero-age main sequence
(ZAMS). The model with only hydrodynamic instabilities has a fast
rotation core. The rotation of model with magnetic fields is,
however, almost uniform. During the early stage of the MS, the
rotation of the model M1.0a is differential, but the model M1.0b is
a quasi-solid body rotation. At the age of 4.5 Gyr, the surface
rotation rates of two models are around 2.7 $\times$ $10^{-6}$
rad/s. However, internal distribution of angular velocity is quite
different. The model M1.0a shows a strong differential rotation with
a factor of about 40 between the angular velocity in the core and at
the surface; but the model M1.0b shows an almost uniform angular
velocity, with a small increase in $\Omega(r)$ in the center of $r$
$<$ 0.2 $R_{\odot}$, as that obtained by Eggenberger et al.
(\cite{egge05}). The surface rotation rate of model M1.0b is higher
than that of model M1.0a in the early evolutionary stage. And the
loss rate of angular momentum is related to $\Omega^{3}$.
Consequently the amount of the angular momentum loss of model M1.0b
is larger than that of model M1.0a. The total angular momentum of
model M1.0a is 2.628 $\times$ $10^{49}$ g $\mathrm{cm^{2}s^{-1}}$ at
the age of 4.5 Gyr, which is about 13 times larger than the
seismical result (1.94 $\pm$ 0.05) $\times$ $10^{48}$ g
$\mathrm{cm^{2}s^{-1}}$ (Komm et al. \cite{kom03}); but the total
angular momentum of model M1.0b is 2.045 $\times$ $10^{48}$ g
$\mathrm{cm^{2}s^{-1}}$ at the age of 4.5 Gyr, which is consistent
with the result of helioseismology at the level of 3 $\sigma$.

   \begin{figure}
     \includegraphics[angle=-90, width=6cm]{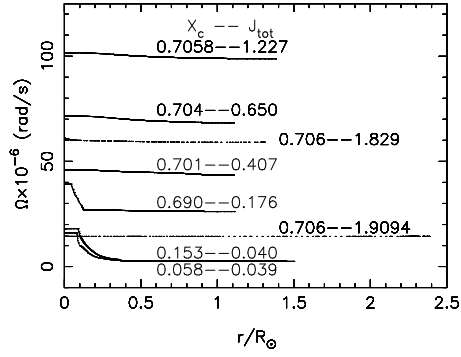}
     \centering
       \caption{Rotation profiles as a function of radius for
       model M1.2 at the different evolutionary stages indicated
       by the central hydrogen $X_{c}$. The initial angular momentum
       $J_{0} =1.9095 \times 10^{50}$ g cm$^{2}$ s$^{-1}$. The $\mathrm{J_{tot}}$
       is the total angular momentum of models in $10^{50}$ g cm$^{2}$ s$^{-1}$.
       The lower dotted line shows the rotation profile of PMS model
       at the age of 1 Myr.
        }
       \label{fig2}
   \end{figure}
Figure \ref{fig2} shows the evolution of the internal rotation
profile of model M1.2. In the early evolutionary stage, the angular
velocity $\Omega (r)$ is almost constant. At the stage of $X_{c}
\sim$ 0.69, about 90 percent of the initial angular momentum has
been lost; the rotation is nearly uniform in the radiative region;
but the rotation of the convective core is faster than that of the
radiative region. There is a transition region between the
convective core and the radiative region, where the angular velocity
has a sharp radial change due to the spin-down of the outer parts of
model resulting from angular momentum loss and the decrease of the
horizontal coupling provided by the magnetic field resulting from
the increase in $\mu$-gradient in the region. The loss of angular
momentum mainly occurs in the early evolutionary stage. During the
late stage of the MS, the total angular momentum of model M1.2 is
only several percent of the initial angular momentum; the rotation
is slow; thus the loss rate of angular momentum is very low.
Consequently, the angular momentum of model M1.2 is almost
conservative from the stage of $X_{c} = 0.153$ to the stage of
$X_{c}=0.058$.

   \begin{figure}
     \includegraphics[angle=-90, width=6cm]{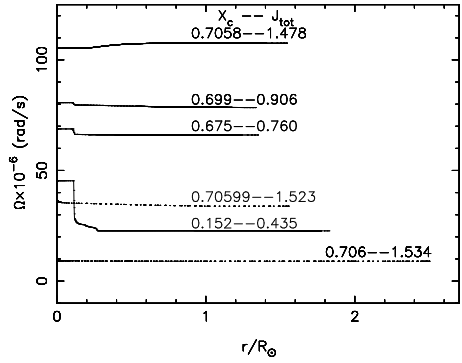}
     \centering
       \caption{Same as Fig. \ref{fig2} but for
       model M1.4. The initial angular momentum
       $J_{0} =1.534 \times 10^{50}$ g cm$^{2}$ s$^{-1}$.
        }
       \label{fig3}
   \end{figure}
   \begin{figure}
     \includegraphics[angle=-90, width=6cm]{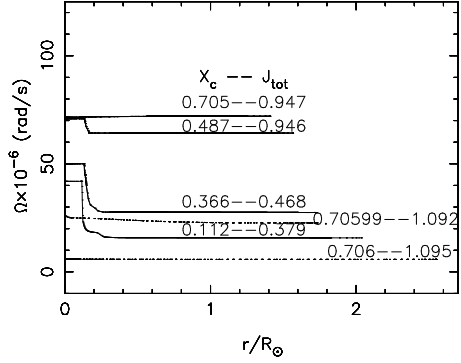}
     \centering
       \caption{Same as Fig. \ref{fig2} but for
       model M1.5. The initial angular momentum
       $J_{0} =1.095 \times 10^{50}$ g cm$^{2}$ s$^{-1}$.
        }
       \label{fig4}
   \end{figure}
The evolution of the internal rotation profiles of models M1.4 and
M1.5 are shown in Figs. \ref{fig3} and \ref{fig4}. The model M1.4
has lost about 50 percent of the initial angular momentum at the
stage of $X_{c} = 0.675$. But the model M1.5 has only lost about 15
percent of the initial angular momentum even at the stage of $X_{c}
= 0.487$. The distributions of the angular velocity of models M1.4
and M1.5 are different from that of model M1.2. The angular velocity
between the convective core and the radiative region of models M1.4
and M1.5 decreases obviously when the radius increases. The radial
change of the angular velocity between the convective core and the
radiative region in models M1.4 and M1.5 is larger than that in
model M1.2, which should be due to the $\mu$-gradient and the fast
spin-down occurring at the same stage in models M1.4 and M1.5.
However, in model M1.2, the fast spin-down occurs in the early
evolutionary stage when the $\mu$-gradient is small.

\begin{figure}
     \includegraphics[angle=-90, width=6cm]{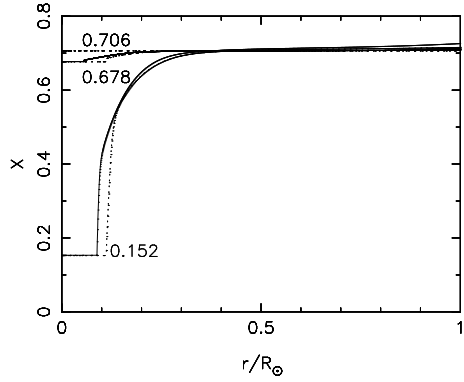}
     \centering
       \caption{Internal distribution of the hydrogen mass fraction
       $X$ as a function of the radius. The solid lines show model
       M1.2. The dotted lines refer to model M1.4.
        }
       \label{fig5}
   \end{figure}

Figure \ref{fig5} shows the distribution of the hydrogen mass
fraction $X$ of models M1.2 and M1.4. It is obvious that there is a
sharp $\mu$-gradient at the bottom of the radiative region of models
M1.2 and M1.4 at the late stage of MS. The $\mu$-gradient and the
$\Omega$-gradient are in the same region. The ratio of magnetic
field, $B_{r}/B_{\phi}$, is related to $\nabla_{\mu}^{-1}$ in
Tayler-Sprut dynamo, namely $D_{m}\sim \nabla_{\mu}^{-2}$. Thus the
increase in $\mu$-gradient must lead to the decrease in the coupling
provided by magnetic fields. This scenario was first found by
Eggenberger et al. (\cite{egge05}) in the Sun.

\section{Discussion and conclusions}

The surface velocity is sensitive to the loss rate of the angular
momentum. The equation (\ref{agloss}) may overestimate the loss rate
of angular momentum of the rapid rotation stars (Andronov et al.
\cite{andr03}). But it can reproduce the Sun's rotation. Thus we
take it in our models. The adjustable parameter, $f_{k}$, is
adjusted to obtain the solar rotation rate at the age of 4.5 Gyr in
models M1.0a and M1.0b. However the same value of $f_{k}$ cannot be
applied to models M1.4 and M1.5 because the convective envelope of
models M1.4 and M1.5 is too shallow. Thus we take a small $f_{k}$
for models M1.4 and M1.5.

The value of parameter $f_{\Omega2}$ is adjusted to obtain a
quasi-solid rotation in our models. The value of 0.01 can do work in
our models. However this value is less than 1. This could be a
consequence of overestimating the ratio of $B_{r}$ to $B_{\phi}$ in
Tayler-Spruit dynamo.

The distribution of angular velocity of models M1.0a and M1.0b shows
that the rotation profile strongly depends on the efficiency of
angular momentum transport. The angular momentum is effectively
transported outward by magnetic fields in M1.0b. Thus the rotation
of the core of M1.0b is slow comparing with that of M1.0a. The
surface rotation rate mainly depends on the loss rate of angular
momentum and the amount of outward transport of angular momentum.
Because both M1.0a and M1.0b have same value of $f_{k}$ and initial
$\Omega$, the discrepancy of the surface velocity between M1.0a and
M1.0b relies on the efficiency of outward transport of angular
momentum. In M1.0b, the loss of angular momentum is counteracted by
magnetic angular momentum transport. Thus the surface velocity of
M1.0b is higher than that of M1.0a when the interior of M1.0b has
enough angular momentum to transport outward. The loss rate of
angular momentum is related to $\Omega^{3}$. Consequently the amount
of angular momentum loss of model M1.0b is larger than that of model
M1.0a. This scenario takes place in the early evolutionary stage.

At the early stage of M1.2, the fast spin-down leads to the sharp
radial change of angular velocity at the top of the convective core.
But at the same stage of models M1.4 and M1.5, the spin-down is
slow. At the late evolutionary stage of M1.2, although there is a
large $\mu$-gradient at the top of the core, the spin-down is very
slow. Thus the radial change of the angular velocity is small
comparing with that of models M1.4 and M1.5 at the top of the core.
However in models M1.4 and M1.5, the $\mu$-gradient and the fast
spin-down resulting from angular momentum loss and stellar expansion
occur on the same stage. Therefore the radial change of angular
velocity is large at the top of the core in models M1.4 and M1.5.

The 1.0 $\mathrm{M}_{\odot}$ model with only hydrodynamic
instabilities has a fast rotation core, and its total angular
momentum is 2.628 $\times$ $10^{49}$ g $\mathrm{cm^{2}s^{-1}}$ at
the age of 4.5 Gyr, which disagrees with the helioseismic results.
However the 1.0 $\mathrm{M}_{\odot}$ model with magnetic fields has
a slow rotation core, and the rotation is almost uniform in the
radiative region, which are consistent with the seismical results.
Moreover the total angular momentum of the model with magnetic
fields is 2.045 $\times$ $10^{48}$ g $\mathrm{cm^{2}s^{-1}}$ at the
age of 4.5 Gyr, which agrees with the helioseismic result at the
level of 3 $\sigma$.

A diffusion coefficient of magnetic angular momentum transport is
obtained. Not only can the magnetic fields reproduce a quasi-solid
rotation, but they can enhance the loss rate of angular momentum.
The rotation of solar-like stars with magnetic fields is almost
uniform in the radiative regions, which is consistent with the
results of helio- and asteroseismology. However there is a
transition region between the convective core and the radiative
region, where the angular velocity has a sharp radial change, which
is different from that of solar model and that of massive stars
shown by Maeder \& Meynet (\cite{maed04}). Moreover the changes of
the angular velocity in the transition region increase with the
increasing in the age and mass.

--------------------------------------

\begin{acknowledgements}

This work was supported by the Ministry of Science and Technology of
the People's republic of China through grant 2007CB815406, and by
the NSFC though grants 10173021, 10433030, 10773003, and 10778601.

\end{acknowledgements}

\label{lastpage}

\end{document}